\documentclass[twocolumn,showpacs,floatfix]{revtex4}

\usepackage{graphicx}
\usepackage{amsmath,amssymb,amstext}
\usepackage{color}

\begin{document}

\title{\textbf{Pseudogap and local pairs in high-T$_c$ cuprate superconductors}}

\author{A.\,L.\,Solovjov\footnote{{}Electronic address:\: solovjov$@$ilt.kharkov.ua} and  M.\,A.\,Tkachenko}

\affiliation{B.\,I.\,Verkin Institute for Low Temperature Physics and Engineering of National Academy of Science of Ukraine, Lenin ave. 47, Kharkov
              61103, Ukraine \\}

\date{\today}

\begin{abstract}
Analysis  of the resistivity  data  recently  reported  by  Kondo et al.\cite {Kon} for  $(Bi,Pb)_2(Sr,La)_2CuO_{6-\delta}$  (Bi2201)
single-crystals has been performed within our model developed to study pseudogap (PG)
in high-$T_c$  superconductors  (HTS's).  The model is  based  on an assumption of the existence of local pairs
in  HTS's at  temperatures  well  above  $T_c$. Comparative analysis of our  results  and  results of ARPES
experiments  reported  by  Kondo et al. suggests  the local pairs  to be  one of  the possible reason of the PG formation.
\end{abstract}

\maketitle

\section{Introduction}
High-T$_c$ superconductors (HTS's) actually possess four main properties \cite {Iye,Max,SP,S}.
First of all it is the high $T_c$ itself, where  $T_c$ is the superconducting transition temperature. The next property is the pseudogap (PG) observed  mostly in underdoped cuprates, for example in $YBa_2Cu_3O_{7-\delta}$ (YBCO) \cite {S}.
Below any representative temperature $T^*\gg T_c$ these high-T$_c$ superconductors transfer into the PG regime which is characterized by many unusual features \cite {TS}.  The other property is the presence of strong electron correlations observed in underdoped cuprates \cite{Sl,Y}.
But these correlations are not found for FeAs-based superconductors \cite{Sad}.
The last but not the least property is the reduced density of charge carriers $n_f$. $n_f$ is zero in antiferromagnetic (AFM) parent state of HTS's and gradually increases with doping \cite{Iye,Max}.
But even in an optimally doped YBCO it is an order of magnitude less than in conventional superconductors (SC's) \cite{Iye,Max,SP,S,TS}.
There is growing evidence that just the reduced density of charge carriers may be a key feature to account for all other properties of HTS's.

Despite a huge amount of papers devoted to the HTS's problem the coupling mechanism resulting in the existence of the Cooper pairs at such high temperatures (for instance $T_c$=134 K for $HgBa_2Ca_2Cu_3O_{8+\delta}$) as well as physical nature of the PG still remain controversial.
However, the point of view where the appearance of a PG in cuprate HTS's is due to formation of paired fermions for $T_c<T<T^*$ gradually gaining predominance \cite{S,St,BL,AZ,MS}.
The possibility of the long-lived pair states formation in HTS's in the PG temperature range was justified theoretically in Refs. \cite{KP,A,Tch,L}.
It has also been shown that paired fermions above $T_c$ have to exist in a form of so-called local pairs \cite{S,KP,L}.
The local pairs are something in between the strongly bound bosons (SBB), which form at high temperatures and satisfy the theory of Bose-Einstein condensation (BEC), and more or less conventional fluctuating Cooper pairs which appear as T approaches $T_c$ and satisfy the Bardeen-Cooper-Schrieffer (BCS) theory \cite{L}.
The point is the SBB may form in the systems with low and reduced $n_f$ only \cite {A,Tch,L,H,ER}.
This condition is realized just in underdoped cuprates (see Ref. \cite{S} and references therein) and in new FeAs-based superconductors \cite{Sad,S1,S2} eventually resulting, in our opinion, in PG appearance.
But, strictly speaking, the presence or absence of PG in FeAs-based HTS's still remains controversial \cite{Sad,M}.

It has been shown theoretically \cite{A,Tch,L,H,ER} that the systems with low and reduced $n_f$ acquire some unusual properties.
In such systems the chemical potential $\mu$ becomes a function of $n_f$, $T$ and of the energy of the bound state of two fermions, $\varepsilon_b\sim \xi(T)^{-1}$.
Here $\xi(T)$ is the coherence length which actually determines the pair size.
Thus, the $\varepsilon_b$ becomes an important physical parameter of a Fermi liquid
and determines a quantitative criterion for dense ($\varepsilon_F\gg |\varepsilon_b|$) or
dilute ($\varepsilon_F \ll |\varepsilon_b|$) Fermi liquid ($\varepsilon_F=E_F$ is the Fermi energy).
It is shown that $k_F\varepsilon_b \gg$ 1 and  $\mu=\varepsilon_F$ in the first case.
On the contrary, $k_F \varepsilon_b \ll$1 and $\mu=-|\varepsilon_b|$/2
($\ne\varepsilon_F$) in the second case.
Besides, the coherence length, $\xi(T)=\xi(0)(T/T_c-1)^{-1/2}$, is extremely short in HTS's \cite{Max,S}, which is an additional requirement for the formation of SBB \cite{L,H,GL}.

As the temperature lowers, $\xi$(T) evidently increases whereas $\varepsilon_b$ decreases.
Thus there must be a transition from SBB to fluctuating Cooper pairs, or from BEC to BCS regime.
The transition was predicted theoretically in Ref. \cite{GL} and experimentally approved in our previous papers \cite{S,S3}.
The fact has to confirm our assumption concerning the existence of local pairs in HTS's. To be more sure, in this paper we have analyzed resistivity data for $(Bi,Pb)_2(Sr,La)_2CuO_{6+\delta}$ (Bi2201) single-crystals recently reported by Kondo et al. in the Supplementary to their ARPES experiments \cite{Kon}.
The analysis was performed in a framework of our local pair model based on the assumption of the existence of local pairs in cuprates \cite{S,S3}.
The measured temperature dependence of the pseudogap is compared with the spectral weight of the energy distribution curves $(EDCs)$ close to the Fermi level, $W(E_F)(T)$, reported in Ref. \cite{Kon}.
\section{Model}
A pseudogap in HTS's is manifested in resistivity measurements as a downturn of the longitudinal resistivity $\rho(T)$ at $T\leq T^*$ from its linear behavior.
This results in the excess conductivity $\sigma'(T) = \sigma(T) - \sigma_N(T)= [1/\rho(T)]- [1/\rho_N (T)]$\, or

\begin{equation}
\sigma'(T)= [\rho_N(T)-\rho(T)]/[\rho(T)\rho_N(T)].
\label{sigma-t}
\end{equation}

Here $\rho_N$(T)=$\alpha$T+b determines the resistivity of a sample in the normal state
extrapolated towards low temperatures.
This way of determining $\rho_N(T)$, which is widely used for calculation of $\sigma'(T)$ in HTS's \cite{S}, has found validation in the Nearly Antiferromagnetic Fermi Liquid (NAFL) model \cite{SP}.

It was found \cite{S,TS} that the conventional fluctuation theories by Aslamasov and Larkin (AL) \cite{AL} and by Maki \cite{Mak} and Thompson \cite{Th} (MT), reanalyzed for the HTS's by Hikami and Larkin (HL) \cite{HL}, well fit the experiment but up to approximately 110 K only.
It is clear to get information about PG we need an equation which describes the whole experimental curve from $T_c$ up to $T^*$ and contains PG in the explicit form.
Besides, the dynamics of pair-creation and pair-breaking above $T_c$ must be taken into account in order to correctly describe the experiment \cite{A,Tch,L,S3,B}.
Unfortunately, so far there is no completed fundamental theory to describe the high-$T_c$ superconductivity as a whole and in particular a pseudogap phenomenon.
Ultimately, the equation for $\sigma'(\varepsilon)$ has been proposed in Ref. \cite{S3} and can be written as

\begin{equation}
\sigma '(\varepsilon) = \frac{e^2\,A_4\,\left(1 -
\frac{T}{T^*}\right)\,\left(exp\left(-\frac{\Delta^*}{T}\right)\right)}{(16\,\hbar\,\xi_c(0)\,
\sqrt{2\,\varepsilon_{c0}^*\,\sinh(2\,\varepsilon\,/\,\varepsilon_{c0}^*})},
\label{sigma-eps}
\end{equation}
where $A_4$ is a numerical factor which has the  meaning of the C-factor in the fluctuation conductivity (FLC) theory \cite{S}.
All other parameters, including the coherence length along the {\it c}-axis, $\xi_c$(0), and the theoretical parameter $\varepsilon^*_{c0}$ \cite{S,S3}, directly come from the experiment.
To find $A_4$ we calculate  $\sigma'(\varepsilon)$ using Eq. (2) and fit the experiment in a range of 3D AL fluctuations near $T_c$  where $ln\sigma'(ln\varepsilon)$ is the linear function of the reduced temperature, $\varepsilon = (T - T_c^{mf})\,/\,T_c^{mf}$ , with a slope $\lambda$ = $-$1/2 \cite{AL}.
$T_c^{mf}$ is a mean-field critical temperature.
Solving Eq. (2) for $\Delta^*(T)$, referred to as a PG, one can obtain \cite{S3}

\begin{equation}
\Delta^*(T) = T\,ln\frac{e^2\,A_4\,(1 - \frac{T}{T^*})}{\sigma '(T)\,16\,\hbar\,\xi_c(0)\,\sqrt{2\,\varepsilon_{c0}^*\,\sinh(2\,\varepsilon\,/\,\varepsilon_{c0}^*)}}.
\label{delta-t}
\end{equation}
Here $\sigma'$(T) is the experimentally measured excess conductivity in the whole temperature interval from $T^*$ down to $T_c^{mf}$.
\section{Results and discussion}
In the framework of our local pair model the PG in YBCO films \cite{S3}, YPrBCO films \cite{S4}, FeAs-based superconductor $SmFeAsO_{0.85}$ with $T_c$ =55 K \cite{S1,S2} and finally in the slightly doped HoBCO single-crystals \cite{S5} was successfully studied.
The basic results have been obtained after the resistivity data for the set of four YBCO
films with different oxygen concentration were analyzed using Eqs. (1-3) \cite{S,S3}.
The films were fabricated at Max Plank Institute (MPI) in Stuttgart by pulse laser deposition method \cite{Hab}.
All samples are well structured {\it c}-oriented epitaxial YBCO films.
It was confirmed by studying the correspondent x-ray and Raman spectra.
The parameters of the films are listed in Table I, where $d_0$ - is the sample thickness. The temperature dependencies of $\Delta^*$ found from our analysis for all four films are shown in Fig. 1.
The main common feature of every plot is a maximum of  $\Delta^*(T)$ observed at the same $T_{max}\approx$ 130 K.
Important in this case is a fact that the coherence length in the {\it ab} plane, $\xi_{ab}(T)$, was found to be similar for all the studied samples: $\xi_{ab}(T_{max})\approx 18\AA$.

\begin{table}[tbp]
\caption [ ] 

The parameters of the YBCO films with different oxygen concentration (sample F1$-$F6).
\centering
\begin{tabular}{||l|c|c|c|c|c|c|c||}
\hline
Sample & $d_0$ & $T_c$ & $T_c^{mf}$ & $\rho(100K)$
& $\rho(300K)$ & T* & $\xi_c(0)$\\ [0.5ex]
 & $(\AA)$ & $(K)$& $(K)$ & $(\mu\Omega cm)$
& $(\mu\Omega cm)$ & $(K)$ & $(\AA)$\\ [0.5ex]
\hline
F1 & 1050 & 87.4 &88.46 & 148 & 476 & 203 & 1.65 \\
\hline
F3 & 850 & 81.4 & 84.55 & 237 & 760 & 213 & 1.75\\
\hline
F4 & 850 & 80.3 & 83.4 & 386 & 1125 & 218 & 1.78\\
\hline
F6 & 650 &  54.2 & 55.88 & 364 & 1460 & 245 & 2.64\\[0.5ex]
\hline
\end{tabular}
\label{tab:sample-values}
\end{table}

Above 130 K $\xi_{ab}$(T) is very small ($\xi_{ab}(T)<18\AA$) whereas the coupling energy $\varepsilon_b$ is very strong.
It is just the condition for the formation of strongly bound bosons (SBB) \cite{Tch,L,H,ER}.
It was found \cite{S3} that in the temperature interval 130 K$ < T < T^*$ every experimental curve of Fig. 1 can be fitted by the Babaev-Kleinert (BK) theory \cite{BK} in the BEC limit (low $n_f$) in which SBB have to exist \cite{A,Tch,L,H,ER,GL}.
The result is believed to confirm the presence of the local pairs in the films which have to exist at high temperatures in the form of SBB.
In accordance with the theory the SBB cannot be destroyed by thermal fluctuations because of very strong bonding.
Besides, they have to be local (i.e. not interacting with one another) objects because the pair size is much less than the distance between the pairs.
As a result, the local pairs demonstrate no superconducting (collective) behavior in this temperature interval.
This conclusion is confirmed, for example, by results of the tunneling experiments \cite{Ya,R} in which the SC tunneling features are smeared out above $T\geq 130 K$.
Thus, in terms of Ref. \cite{Kon}, above $T_{max}$ we have the so-called nonsuperconducting part of the pseudogap \cite{S,S3}.

\begin{figure}[t]
\begin{center}
\includegraphics[width=.45\textwidth]{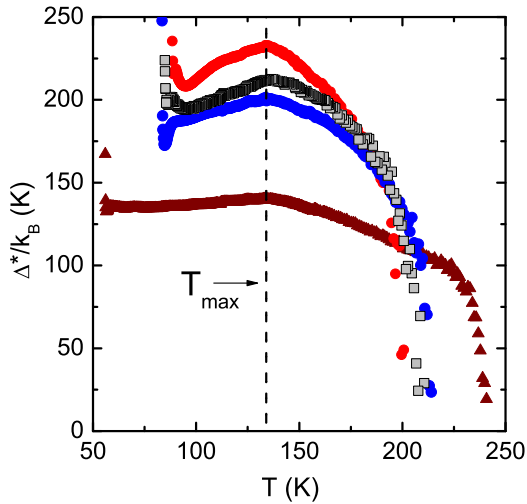}
\caption{Dependence of $\Delta^*/k_B$ on T calculated by Eq. (3) for samples F1 (upper curve, red dots), F3 (gray squares), F4 (blue dots) and F6 (low curve, triangles). $T_{max} \approx 130 K$.}
\end{center}
\end{figure}

With decrease of temperature $\xi_{ab}(T)$ continues to increase whereas $\varepsilon_b(T)$  is getting smaller.
Ultimately, at $T\leq T_{max}\approx 130 K$ the local pairs overlap and acquire the possibility to interact.
Besides, they can be destroyed by thermal fluctuations now, i.e. transfer into fluctuating Cooper pairs.
The superconducting (collective) behavior of the local pairs in this temperature region is distinctly seen in many experiments.
These are again the tunneling measurements \cite{Ya,R} in which the SC tunneling features are well observable now.
In Ref. \cite{KT} the currents of coherent bosons with charge 2e were observed up to $T\sim$ 130 K when studying the quantization of the magnetic flux on YBCO films. Eventually, the direct imaging of the local pair SC clusters persistence up to approximately 130 K in optimally doped Bi2212 is recently reported in Ref. \cite{Y}.
Thus, below $T_{max}$ this is the superconducting part of the pseudogap \cite{S,S3}.

Moreover, $\xi_{ab}(T_{max})\approx 18\AA$ is considered to be the critical size of the local pair in YBCO systems \cite{S,S3}.
When $\xi_{ab}(T)<18\AA$ the local pairs behave like SBB and transfer into fluctuating Cooper pairs when $\xi_{ab}(T)>18\AA$  below $T_{max}$.
Thus, the BEC-BCS transition is experimentally observed in our experiments as illustrates Fig. 1.
Eventually we may conclude that the PG description in terms of local pairs gives a set of reasonable and self-consistent results.
However, to justify the conclusion it would be appropriate to have independent results of other research groups, who have measured straightforwardly the PG or any of its components.

Fortunately analysis of the pseudogap in optimally doped $Bi_2Sr_2CaCu_2O_{8+\delta}$ (Bi2212) single-crystals with $T_c$=90K (OP90K) and in $(Bi,Pb)_2(Sr,La)_2CuO_{6+\delta}$ (Bi2201) single-crystals with various $T_c$'s by means of ARPES spectra study was recently reported \cite{Kon}.
The studies were mainly performed on Bi2201 samples because Bi2201 has a low $T_c$ of ~35 K, but large $T^*$ similar to that of Bi2212 at optimal doping.
This makes it possible to investigate separately characteristics of the superconducting and pseudogap states.
Besides, Bi2201 was chosen to avoid the complications resulting from the bilayer splitting and strong antinodal bosonic mode coupling inherent to Bi2212 \cite{Kon,RH,NS}.

The temperature evolution of the spectral line shape, measured at the antinodal Fermi momentum was finally studued \cite{Kon}.
Symmetrized EDCs demonstrate the opening of the pseudogap on cooling below $T^*$.
Here $T^*$ was defined as a temperature at which the two peaks in symmetrized EDC merge into one peak at increased temperature.
The temperature dependence of the resistivity along the $\it ab$ plane, $\rho_{ab}$, was also measured.
From the resistivity measurements $T^*$ was determined in the usual way as a downturn of  $\rho_{ab}(T)$ at $T^*$ from its linear dependence at high temperatures.
It was found that $T^*$ obtained from the resistivity measurements agrees well with one determined from the ARPES data using "single spectral peak criterion" \cite{Kon}.

Finally from the ARPES experiments information about temperature dependence of the loss of the spectral weight close to the Fermi level, $W(E_F)$, was derived \cite{Kon}.
The temperature dependence of $W(E_F)$ for OP35K Bi2201 ($T_c$=35 K, $T^*$ = 160 K) is shown in Fig. 2a taken from Ref. \cite{Kon} and turned out to be essentially unusual. Above $T^*$ the $W(E_F)$ is nonlinear function of $T$, and starts to decrease after reaching the maximum around 210 K.
This is an expected behavior for metallic samples, where thermal effects became dominant and broaden the spectral peak at $E_F$, causing the peak height to decrease \cite{Kon}. But below $T^*$, in the temperature interval $T^*>T>T_{pair} = (100 \pm 5)$ K (Fig. 2a), the spectral weight decreases linearly, which is considered as a characteristic behavior of the "proper" PG state \cite{Kon}.
However, no assumption as for physical nature of this linearity as well as for existence of the paired fermions in this pseudogap region is proposed.
Below $T_{pair}$ the $W(E_F)$ vs $T$ noticeably deviates from the linearity towards lower values.
The deviation suggests the onset of another state of the system.
In accordance with Ref. \cite{Kon}, this state probably arises from the electron pairing because the weight loss associated with this state smoothly evolves through $T_c$ (Fig. 2a).

To compare results and justify our local pair model, the $\rho_{ab}$ vs $T$ curve of OP35K Bi2201 reported in the Supplementary to Ref. \cite{Kon} was studied.
The data were analyzed within our model.
Resulting $\Delta^*(T)/\Delta^*_{max}$ is plotted in Fig. 2b (green dots).
The $\Delta^*(T)$ jumps in the curve are most likely due to changes of the measuring rate during the experiment.
The $\Delta^*(T)$ was calculated by Eq. (3) with the following reasonable set of parameters: $T_c$ = 35 K, $T_c^{mf}$ = 36,7 K, $T^*$ =160 K, $\xi_ñ(0)\approx 1.5\AA$, $\xi_{c0}^*$ = 0.885, and  $A_4$ = 25.
$\sigma'(T)$ is the experimentally measured excess conductivity derived from the resistivity data using Eq. (1).

As expected the shape of the $\Delta^*(T)$ curve is similar to that found for YBCO films (see Fig. 1).
As it is seen in the figure the maximum of $\Delta^*(T)/\Delta^*_{max}$  at $T_{max}\approx 100 K$  actually coincides with $T_{pair}$ which seems to be reasonable.
Really, in accordance with our logic $T_{max}$ is just the temperature which divides the PG region on SC and non-SC parts depending on local pair state as described above.
Above $T_{max}$ the local pairs are expected to be in the form of SBB.
Most likely just the specific properties of SBB provide the linear W($E_F$) vs $T$ dependence in this temperature region (Fig. 2a).
The two following facts are believed to confirm the conclusion. Firstly, when SBB disappear above $T^*$, the linearity disappears too.
Secondly, below $T_{max}$, or below $T_{pair}$  in terms of Ref. \cite{Kon}, the SBB transform into fluctuating Cooper pairs giving rise to the collective (superconducting) properties of the system.
This argumentation coincides with the conclusion of Ref. \cite{Kon} as for superconducting part of pseudogap below $T_{pair}$. As SBB are now also absent the linearity of W($E_F$) vs $T$ again disappears.
Thus, $\Delta^*$(T) found from our analysis (Fig. 2b) is in good agreement with the temperature dependence of the loss of the spectral weight W($E_F$) (Fig. 2a) obtained from the ARPES experiments performed on the same sample.
In this way, the results of ARPES experiments reported in Ref. \cite{Kon} are believed to confirm our conclusion as for the existence of local pairs in HTS's, at least in Bi2201 systems.

\begin{figure}[t]
\begin{center}
\includegraphics[width=.43\textwidth]{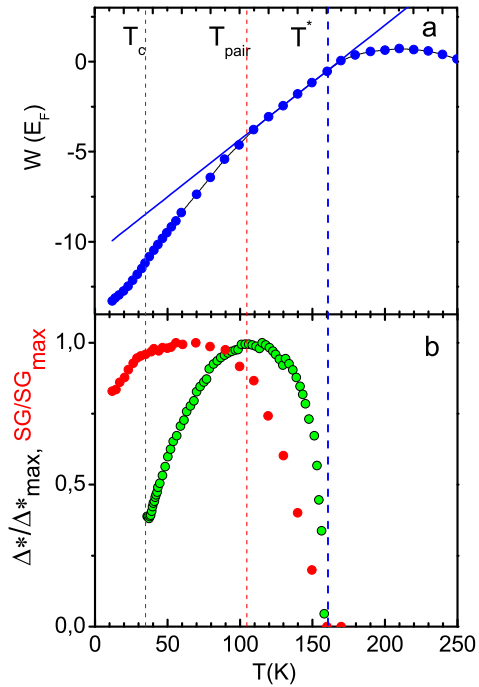}
\caption{a. Spectral weight $W(E_F)$ vs $T$ (blue dots) for  OP35K Bi2201. The solid line is the guidance for eyes only. b. Pseudogap $\Delta^*/\Delta_{max}$ (green dots) and spectral gap $SG/SG_{max}$ (red dots) as the functions of temperature for the same sample.
}
\end{center}
\end{figure}

The OP35K Bi2201 was chosen to be analyzed because it is the only sample for which the spectral gap ($SG$) equals to the energy of the spectral peaks of EDC was defined as a function of temperature \cite{Kon}.
The $SG(T)$ divided on its maximum value, $SG_{max}$, is also plotted in Fig. 2b (red dots).
At first sight there is a qualitative agreement between both curves shown in Fig. 2b. However, there are at least two differences.
The first one is the evident absence of the direct correlation between the $SG(T)$ and the temperature dependence of the loss of the spectral weight $W(E_F$) (Fig. 2 a, b).
Why the maximum of $SG$ is shifted towards low temperatures in comparison with $T_{pair}$ (Fig. 2 a, b) has yet to be understood.
The second difference is the absolute values of $SG$ and $\Delta^*$.
The spectral gap has $SG_{max} \approx$ 40 meV and $SG(T_c) \approx$ 38 meV \cite{Kon}.
It gives $2SG(T_c)/k_BT_c \approx$ 26 which is evidently too high. Our pseudogap values are $\Delta^*_{max} \approx$ 16.5 meV and $\Delta^*(T_c) \approx$ 6.96 meV, respectively. It gives $2\Delta^*(T_c)/k_BT_c \approx$ 4.57 which is a value of common occurrence for HTS's \cite{Max,S,TS}.
Both differences suggest an issue that clear correlation between SG and PG still remains controversial.
\section{Conclusion}
In conclusion, analysis of $\rho_{ab}$ vs $T$ curve of OP35K Bi2201 \cite{Kon} performed within our local pair model has evidently shown that found $\Delta^*(T)$ is in a good agreement with the temperature dependence of the loss of the spectral weight W($E_F$) measured by ARPES for the same sample.
Besides, the values of  $\Delta^*(T_c) \approx$ 6.96 meV and $2\Delta^*(T_c)/k_BT_c \approx$ 4.57 are common for HTS's \cite{S}.
It allows us to reasonably explain the W($E_F$) vs $T$ dependence both above and below $T_{pair}$ in terms of local pairs.

The obtained results are also in agreement with the conclusion of Ref's. \cite{Y,RH,NS} as for SC and non-SC parts of the PG in Bi systems.
Besides, the formation of local pairs is believed to explain the rise of the polar Kerr effect (PKE) and response of the time-resolved reflectivity (TRR) both observed just below $T^*$ \cite{RH}.
Whereas, the Nernst effect \cite{Y}, which is likely due to the superconducting properties of the local pairs, is observed only below $T_{pair}$, or $T<T_{max}$ in terms of our model.



\end{document}